# Engaging and Educating Eclipse Observers Through Workshops, Media, Planetarium shows and Citizen Science

Patricia H. Reiff[1], Carolyn T. Sumners[1,2], Charles H. Gardner[1], Amir Caspi[3], and Sarah Kovac[3,4]

1. Department of Physics and Astronomy, Rice University; 2. Houston Museum of Natural Science; 3. Southwest Research Institute; 4. High Altitude Observatory, National Science Foundation National Center for Atmospheric Physics

## 1. Abstract

The "Heliophysics Big Year" was an extended "year" where major solar events engaged the public. It included two eclipses (annular on October 14, 2023; and total on April 8, 2024), plus solar max, plus the closest approach of Parker Solar Probe to the Sun on December 24, 2024. After the eclipse of 2017, many millions more Americans looked forward to observing the solar corona for only the first (or second) time. We expanded our eclipse website https://space.rice.edu/eclipse/ with activities, citizen science projects, resources, downloadable Powerpoints, training videos, suggested equipment, and links to other compendia. We worked with the Citizen CATE 2024 project as Southwest Regional Coordinator, training the state coordinators and their teams with the specialized equipment and procedures. We trained teachers at local, regional (CAST) national (NSTA), and international (SEEC) workshops, providing eclipse viewing cards, lenses to make "solar cup projectors," a pattern for a safe viewing screen, and access to the training materials. We made presentations to the media at SciLine in San Antonio, and hosted public events to demonstrate safe eclipse viewing techniques. HMNS also hosted live viewing for both the annular and total plus solstice and equinox events, reaching tens of thousands of people. HMNS also secured from Buc-ees a grant to provide 100 eclipse viewing cards for every public school (8,800+) in Texas. We distributed another 57,000 eclipse viewers to teachers and the public in our events. We appeared in media both in advance of the eclipses and as live commentators. The most lasting and impactful product was our planetarium show "Totality," which was given away free and shown in various formats (flatscreen, fisheye, or prewarped). Over 180,000 views of the show and its animations have been documented. Unfortunately, our totality viewing site was clouded out, but most of the CATE 2024 sites were able to acquire data. We continued to improve our space weather forecasting site https://mms.rice.edu. We used our email lists (14,000+) to send out a real-time warning of the major solar storms of May 10–11 and October 8–10, 2024. In total, we provided nearly two million people with Heliophysics information.

## 2. Introduction

Rice University and the Houston Museum of Natural Science (HMNS) have had an ongoing formal outreach association in space science since 1988, when we created the first NSF-funded "Teacher Research Program" in Astronomy and Earth Science. This credit-bearing program brought teachers for a summer to work alongside faculty and graduate students in real research and at the same time creating curriculum that could be brought back to their home schools. This success led to the "Master of Science Teaching" program at Rice (https://mst.rice.edu).

The collaboration between Dr. Carolyn Sumners, Vice President for Astronomy and Physical Science at HMNS, and Prof. Patricia Reiff, Physics & Astronomy, led to many other HMNS/Rice "Firsts" in educational technology, including the world's first internet-connected science kiosk, the US's first video planetarium, the first Earth Science planetarium show, and the first video portable planetarium (https://space.rice.edu/rice_hmns_firsts.html), primarily funded by NASA Cooperative Agreements [*Sumners and Reiff*, 2004]. Our research showed that learning in a dome leads to greater content retention than viewing on a flatscreen [*Zimmerman et al.,* 2014].

Reiff is a Heliospheric physicist and Co-Investigator of the Magnetospheric Multiscale Mission (MMS), and SME expert for a number of outreach programs, including NASA HEAT (Heliophysics Education Activation Team) and its predecessors. She is also an avid eclipse chaser, having led many eclipse expeditions (18 totals and 3 annulars as of this writing).

For the total eclipse crossing the US in 2017, we created, with NASA funding, several solar and lunar eclipse animations https://space.rice.edu/eclipse/eclipse_animations.html that could be used in a classroom or in a planetarium, documented in [*Reiff and Sumners*, 2017]. These were distributed free to planetariums and educators, and used by the media. There was a voice track that could be used if desired. We did not keep an archive of the requests at that time but dozens of planetariums, the media, and educators downloaded the animations, and we used the animations in our workshops for teachers for the 2017 eclipse. One of our online eclipse trainings was posted on YouTube: https://www.youtube.com/watch?v=0ju1WV5O7rY

## 3. Materials and Methods

## Heliophysics "Big Year" (HBY) Activities

For the Heliophysics "Big Year" (2023–2024) we put our outreach efforts into high gear, attempting to reach as many of the public as possible with information about eclipse science and eclipse safety. For this we targeted primarily teachers, because of the multiplicative effect, but also used the media, websites, email lists, and most importantly our planetarium show "Totality." In addition, Reiff was the southwest regional director for the Citizen CATE 2024 project, a participatory science ("citizen science") program led by Dr. Amir Caspi at the Southwest Research Institute (SwRI) in Boulder, CO to observe the eclipse from 35+ identical sites along the eclipse path.

**Updated Eclipse Animations**

In 2021, we added three more eclipse animations: the geometry and two close-up views of an annular eclipse. We began distribution of those in October 2021 and began keeping track of the animation users. Since late 2021, 186 educators (164 planetariums and 22 schools) have requested the animations, in 33 countries and 33 US states (yellow tags in the map below). We also posted the animations in YouTube and got over 19,000 views. We will post the opportunity again as the 2026 and 2027 eclipses approach (which will unfortunately not cross the Americas).

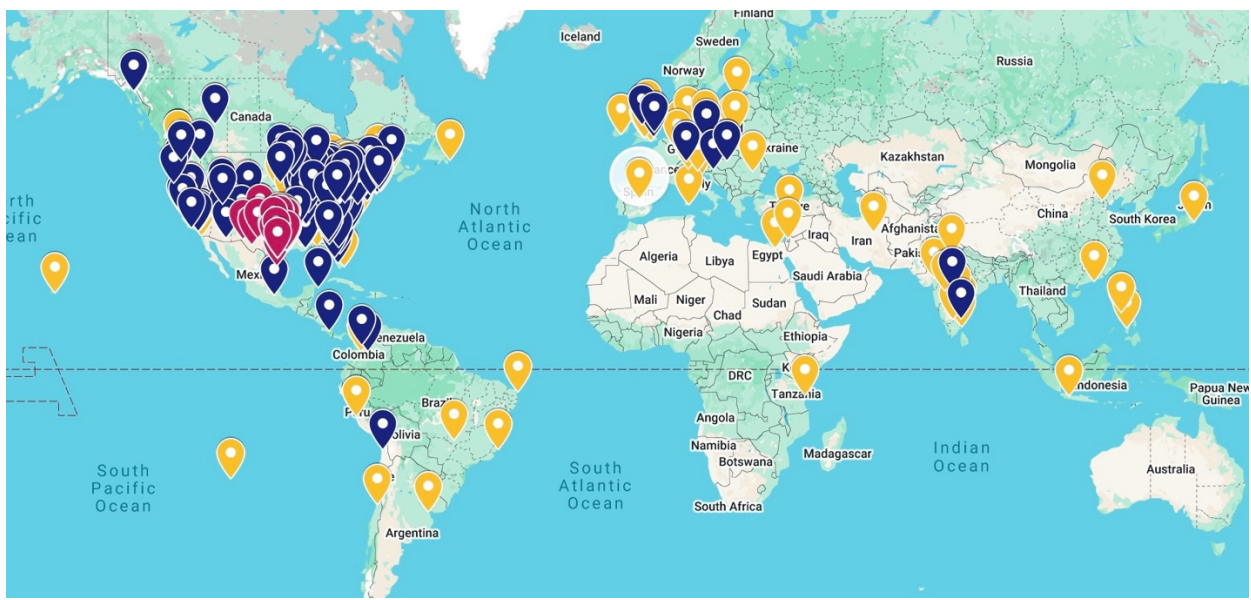

*Fig. 1. World map of the planetarium and school downloads of our eclipse animations (yellow tags), "Totality" show (blue tags) and special Texas version "Totality Over Texas" show (red tags).*

One of the animations that Don Davis created for us had an unusual confirmation. We noted in our animation, "Lunar eclipse from the Moon" (https://www.youtube.com/watch?v=dEEHtstD7Rg), that if you were on the Moon during a total lunar eclipse as seen from Earth, it would be a total solar eclipse for you. However, the Earth would be four times the diameter of the Sun, meaning that the corona would only be seen at the beginning of the eclipse on one side of the Earth and at the end of the eclipse on the other side. In between, the Earth's atmosphere would glow red from the scattered sunlight that gives the full Moon its dark red color. Then, in March 2025, the "diamond ring" from the total lunar eclipse was viewed by the "Blue Ghost" spacecraft placed on the lunar surface by Firefly Aerospace.   Figure 2a (left) shows our prediction from the animation. Figure 2b (right) shows the Blue Ghost image. Of course, the animation has a better effective dynamic range than the lander imager.

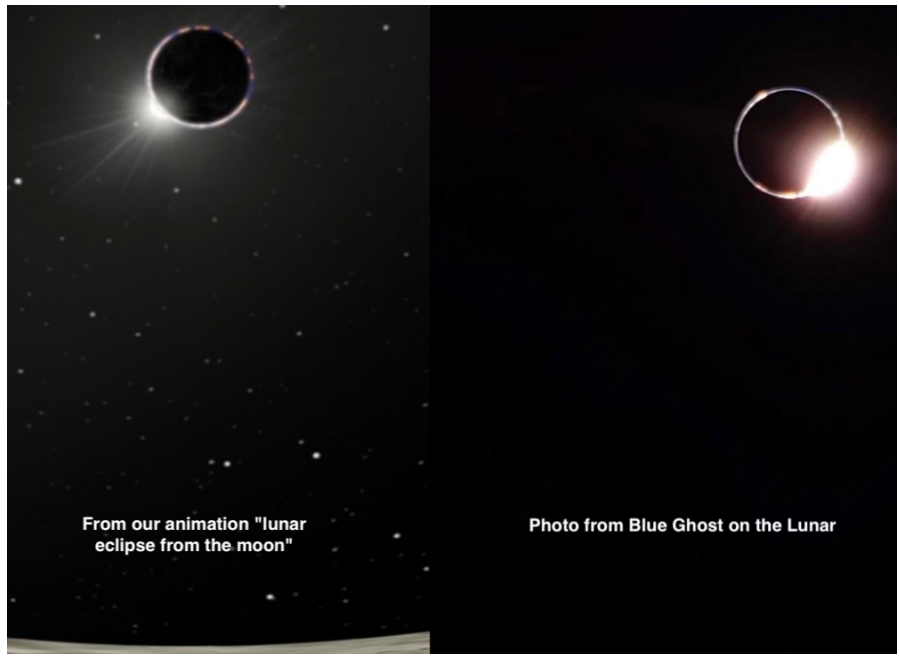

*Fig 2a (left). Our predicted "diamond ring" as seen from the Moon (from our "Lunar Eclipse from the Moon" animation). 2b (right): Diamond ring observed by the Blue Ghost on the lunar surface (posted on X by Firefly Aerospace, 3/14/2025).*

**Planetarium Show "Totality"**

<u>The Show:</u>  The capstone of our effort was the creation of a fulldome planetarium show "Totality!". The show, which used some of our NASA-supported animations, was a cooperative project between HMNS, Rice, and ePlanetarium (https://www.eplanetarium.com).  HMNS provided imagery, additional simulations created by using OpenSpace, script and production, and of course Rice provided the animations and content review. ePlanetarium provided expenses for the narration, formatting, and distribution.

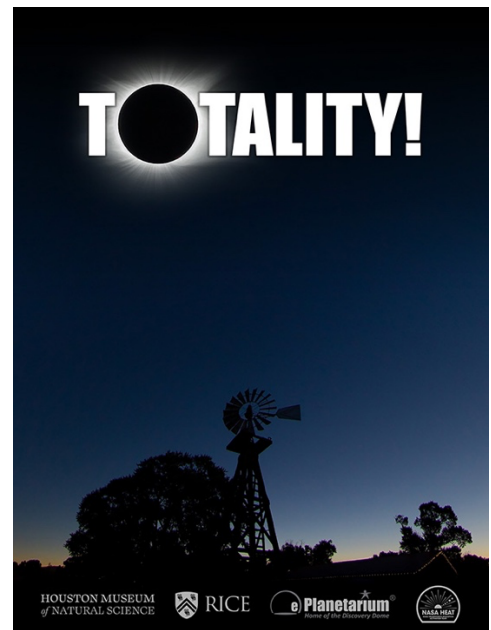

*Fig 3. "Totality" Poster*

The show includes a section on the native Mexican eclipse lore and the 1991 eclipse, the history of eclipse observations, the dynamics and geometry of solar eclipses, the reason why we don't get an eclipse every month, and the sky which will be visible for the April 2024 eclipse.  It then shows the paths of the 1991 and 2009 eclipses and discusses Saros cycles and the next eclipses in that cycle, 2027 in Egypt and 2045 in the US. It also discusses "eclipses" from Mars (which are really just transits) and from Jupiter (which are really just "occultations"). There is then an animation on how the Moon may have formed. Next is an animation of the phases of the eclipse and a discussion of the history of eclipse observations, including the last total eclipse which crossed Texas in 1878; the eclipse observations which proved General Relativity in 1919, and the Antikythera Mechanism. Finally, the show presents images from the 2017 eclipse in Wyoming, including a discussion of safe eclipse viewing techniques.

The movie then shows the path of the annular eclipse of October 14, 2023 and shows the geometry and phases of annular eclipses. It then shows the detailed path of the April 8, 2024 total eclipse shadow. The movie ends with the future of solar eclipses as the Moon slowly drifts away from the Earth, leaving only annulars in the distant future.

The show was distributed free of license fees in several formats: flatscreen for schools and the media, fisheye 2K and 4K versions for small and large planetariums, and in a "prewarped" format for mirror-based planetariums. The users only had to pay a sliding scale format fee. The users had to agree to provide metrics, and just over half did so. The metrics shown below are thus lower limits to the actual visitors shown.

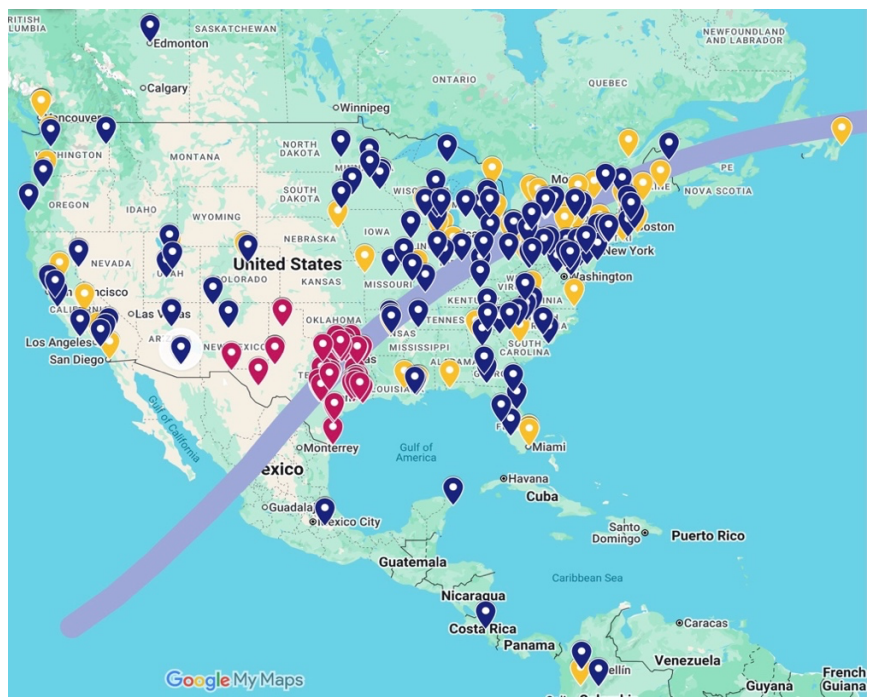

*Fig 4. Planetarium distribution map for "Totality" (blue), "Totality over Texas" (red) and eclipse animations (yellow). The path of totality is shown in light blue.*

Distribution Metrics: The show "Totality" was distributed to 147 planetariums. The "Totality over Texas" version was distributed to 53 Texas planetariums and schools. All of the requests got English versions, with optional English subtitles. In addition, 34 sites got Spanish audio tracks and an additional 16 got Spanish subtitles to show with the English version. One site created a Romanian audio track.

Viewing Metrics:  Of the 200 sites which received copies of the show, we received metric reports from 113 of them.  We had requested separate metrics for showings before and after the annular, and 28 of the sites reported showings before and after the annular eclipse.

The total views from the reporting institutions were 133,469, with an additional 8,800 from HMNS (who offered it daily for more than a year).  In addition, the show was available from FullDomeOnDemand.com, with 278 planetariums renting the show for three days.  YouTube views topped 5,000, including Spanish and Romanian.  Thus, the total show impact is well more than 180,000, not counting the 87 sites which did not report metrics.

**Teacher Programs**

We offered both formal and informal teacher education programs.  In formal education, Rice University offers four semester-long courses for teachers. Teachers could either take the courses for three hours of graduate credit each, or just for 45 hours of Professional Development. These courses are: ASTR502, Solar System for Teachers, https://mst.rice.edu/ASTR502; ASTR503 , Astronomy for Teachers https://mst.rice.edu/ASTR503; ASTR530, Teaching Astronomy Lab https://mst.rice.edu/ASTR530; and PHYS 501, Physics of Ham Radio, https://mst.rice.edu/PHYS501.  We emphasized eclipse science and eclipse safety in all four of the classes starting before the 2017 eclipse but continuing through the 2023/2024 eclipse season.

In addition, we presented many teacher workshops on eclipse science and eclipse safety. In each of these, we gave away eclipse viewers (cards and/or eclipse glasses).  In the later workshops, we also gave away 1.0 diopter lenses which could be used to make our "eclipse solar projector."

**Table 1.  Educators Reached**

| DATE | EVENT | LOCATION | NUMBER OF EDUCATORS |
|---|---|---|---|
| **1/27–28/2022** | Eclipse teacher trainings | Commonwealth Charter Academy, Pittsburgh & Harrisburg PA | 11 |
| **SPRING 2022** | ASTR 503 | Astronomy for Teachers Class, Rice U | 11 |
| **3/31–4/2/2022** | Eclipse Workshop at NSTA | Houston | 448 |
| **6/21/2022** | TLIIST STEM academy | Eclipse training to teachers | 19 |
| **7/21–23/2022** | NSTA, Chicago | Eclipse training in the Discovery Dome | 345 |
| **FALL 2022** | PHYS 502 | "Physics of Ham Radio" course, Rice U | 12 |
| 11/10–12/2022 | "Eclipses 2023-24: Where to go, how to observe safely" | Workshop and booth at CAST science teacher conference, Dallas | 533 |
| 2/10/2023 | Eclipse teacher workshop | SEEC, UHCL, Clear Lake | 25 |
| 3/8/2023 | Eclipse presentation | ESIP online | 22 |
| 5/22/2023 | Eclipse Presentation | SEPA, Kingsport, TN | 120 |
| 7/19/2023 | Eclipse workshop for teachers | ESIP summer workshop, Vermont | 24 |
| **FALL 2023** | ASTR 502 | Teaching Earth and Space Science, RiceU | 8 |
| 8/9/2023 | Eclipse training | Turlington Elementary School | 125 |
| 9/13/2023 | "Eclipses 2023-24: Where to go, how to observe safely" | Presentation at LIPS conference, Detroit (included Totality show) | 76 |
| 10/5/2023 | Eclipse presentation | UHCL STEM luncheon | 135 |
| 10/9–10/2023 | National Parks Event | Booth in conjunction with AAS, Dallas Convention Center | 280 |
| 10/14/2023 | Annular Eclipse Observing | Corpus Christi, TX plus CATE teams | 48 |
| 11/4/2023 | Eclipse Workshop | Wilson Symposium, Houston Christian U | 20 |
| 11/9–11/2023 | Dome showing "Totality", eclipse glasses and maps | CAST (science teachers of Texas conference), Houston | 927 |
| 11/30/2023 | Dome showing "Totality", giving out eclipse glasses | NASA day at the Buffalo Soldiers Museum | 126 |
| 12/14/2023 | Eclipse outreach poster, science talks | AGU meeting, San Francisco | 112 |
| **SPRING 2024** | ASTR 503 | "Astronomy for Teachers", Rice U | **11** |
| 1/20/2024 | CATE training for state coordinators | SWRI, San Antonio | **23** |
| 1/26–27/2024 | CATE training for south Texas teams | Community College, Uvalde | 18 |
| 2/3–4/2024 | CATE training for north Texas teams | Frontiers of Flight Museum, Dallas | 20 |
| 2/9/2024 | Eclipse Panel at SEEC conference | Space Center Houston | 75 |
| 2/10/2024 | Eclipse Workshop at SEEC | Space Center Houston | 65 |
| 2/29/2024 | Eclipse talk ("Out to Lunch") | ESIP via Zoom | 35 |
| 3/2/2024 | Eclipse video at STEM event | UHCL, Clear Lake | 120 |

| 3/16/2024 | ESIP Educators Workshop | Via Zoom, Rice University | 85 |
|---|---|---|---|
| 3/21–23/2024 | Eclipse Planetarium Show and materials | NSTA Conference, Denver Convention Center | 750 |
| 4/2/2024 | "Tips and Tricks for Observing the Eclipse" | Live via Zoom, UC Berkeley | 75 |
| 11/14–16/2024 | Solar Max presentations | CAST conference, San Antonio | 400 |
| **TOTAL FOR TEACHERS** | | In person or in live interactive zoom | **5104** |

Note that in the table above, we did not count events such as colloquia, public events, or student trainings where teachers were present, but were not the prime focus. Those events added several thousand live in person and many additional via webcasts and media events.

**Website:**

One of our prime routes of disseminating information is our "eclipse" website https://space.rice.edu/eclipse. We have been collecting eclipse information, images and information about eclipses since 2004, and have been actively updating our eclipse site since 2014, with major additions for the 2017 and 2024 eclipses. The website is frequently used and is a major source for new subscribers to our eclipse newsletter, which peaked at 1,842 users (in addition to the

*Fig 5. The "eclipse graphics" page on our website. In the remaining figures, the sidebar and headers are not shown.*

12,000 teachers in our teacher "mailman" and 2,200 in our teacher "mailchimp" list).

The key sections of the site are:

- Citizen CATE 2024 experiment
- Eclipse Activities
- Eclipse Animation
- Eclipse Equipment
- Eclipse Graphics
- Eclipse Media Coverage
- Eclipse Resources
- Eclipse Show
- Eclipse Timeline
- Eclipse Training
- Lunar Eclipse FAQs
- Solar Eclipse FAQs
- Safe Eclipse Observing
- Eclipse TicTacToe Game

Each of these sections included links to download materials we developed or links to other sites.

Although our website did not register "hits," we did send a request for feedback to the people on our "eclipse" email list.

Here are the feedback results:

The average number of people who received information from EACH of the respondents was over 500, indicating a strong multiplicative effect. The section of the website that was used most often with students was the Animations (75%) and Safe Observing Techniques (66.7%), with activities, FAQs and graphics all tied at 42%. The sections that were judged the most useful were Animations, Graphics, Activities, and the Eclipse Movie.

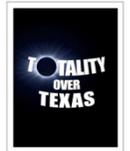
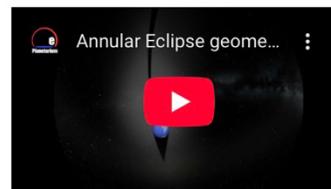
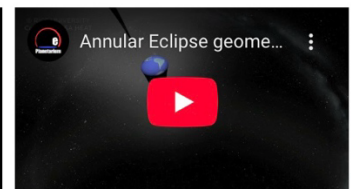
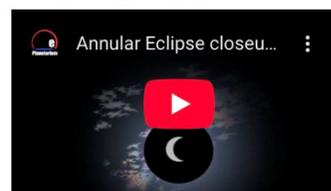
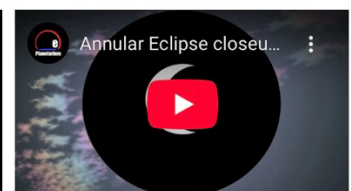
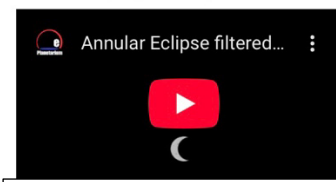
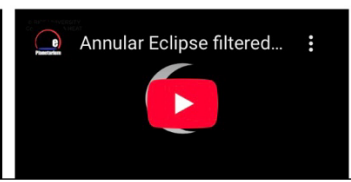

*Fig 6. Eclipse animations page with links to the lunar, solar and annular animations. All are available free in fisheye, flatscreen, and prewarped formats.*

Here are some quotes from the feedback:

 "As the STEM/Science Coordinator at my school, I utilized and shared these resources with all of the teachers and classes at our PreK–8th grade school. We had an eclipse watching party on the school grounds during the last solar eclipse. Your materials were SO helpful! We purchased eclipse glasses for the entire school. For the younger students (PreK–2nd), we used the paper plate "mask" with the glasses to provide further protection for our younger viewers."

"Also used the Solar Eclipse animation in a small exhibit set up at the public viewing site for the actual eclipse – attended by approximately 4,000."

"Thank you, Dr. Pat! Our entire school (Grades K–8) had an eclipse party using your activities and resources from the SEEC presentation. I appreciate your help!"

**Activities:**

The "Activities" page assembled some eclipse activities suitable for learners of all ages.  We included a link to the NASA Citizen Science page for more options, and included several easy-to-follow writeups.

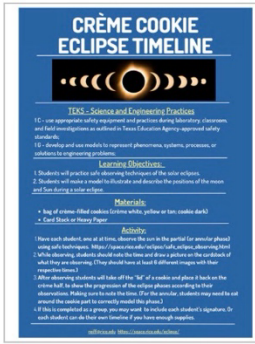
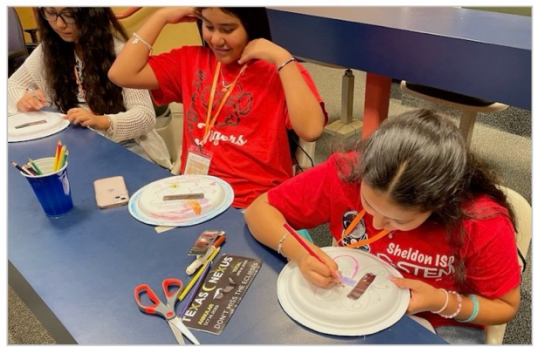
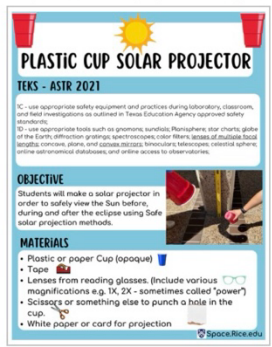

*Fig 7.  Eclipse activities page.  Some were adapted, but the solar cup projector was a new item we created.*

The activity handouts included:

- Crème cookie eclipse timeline
- Paper plate mask (to use with eclipse glasses or viewing cards)
- Solar Cup Projector – easy way to make a brighter, larger image than a pinhole.
- Monitoring Animal Behavior sheet
- Monitoring Weather data sheet

The solar cup projector is an outgrowth of our design for the solar projector on the HMNS sundial (https://space.rice.edu/sundial/pdf/HMNS_Sundial_Info.pdf). That sundial has a special silver ball on the top of the gnomon with three special holes to allow the sun to cast an image at solar noon on the equinoxes and solstices. With a large hole and a low-diopter focusing lens, a bright image can be brought into focus.

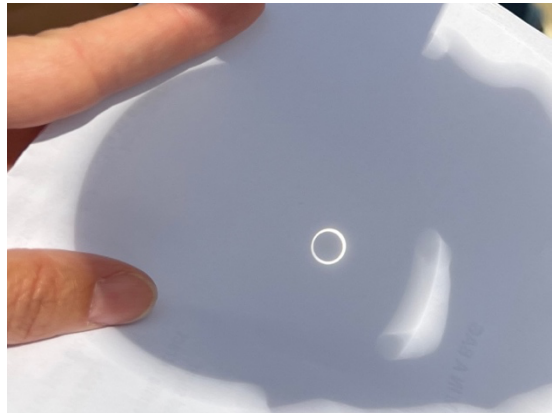

*Fig 8. Solar cup projector casting an image of the annular eclipse, Corpus Christi, TX, October 14, 2023. Courtesy Deborah Edwards.*

The solar cup projectors use an opaque plastic cup with a hole cut in the bottom and a low-diopter lens taped over the hole. A 1.5× (1.5 diopter) lens (which can be found in reading glasses from a dollar store), will focus a solar image at 67 cm. A 1× lens focuses a larger (1 cm) image at 1 m; and a 0.5× lens (which we obtained from an optician) focuses a 2 cm image at 2 m.

The "monitoring animal behavior" sheet and "monitoring weather data" sheet allows students to make observations and record their findings without looking at the sun.

**Safe Solar Observing:**

Our safe solar observing page had links to safe observing compendia by NASA, AAS, and others. It also had patterns for our "PREPARE" poster and our "Safe Solar Screen." It had a link to our safety guide and also the guide in HTML. It also had a link to our 20-minute safety training video, which demonstrated the safe use of several kinds of equipment that we recommend, https://www.youtube.com/watch?v=NXyUEXIFruE.

One special piece of equipment we designed was for teachers who were not allowed to take their students outside. This special "solar screen" was designed to hang over a doorway with cutouts to tape on solar viewers so that the students could look at the sun only through the viewers. The pattern was posted on our website for free download.

We distributed thousands of flyers with the eclipse paths and links to safe observing sites, and created special graphics for the million eclipse viewers. (Figures 10, 11)

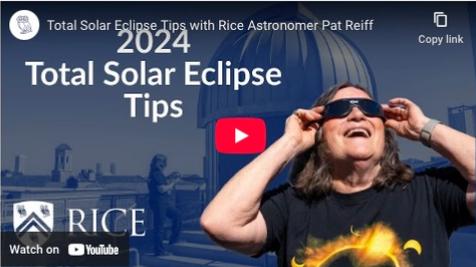

*Fig 9. Eclipse safety page*

*Fig 10. Eclipse map of the "Texas Nexus", with thousands distributed with the eclipse viewers.*

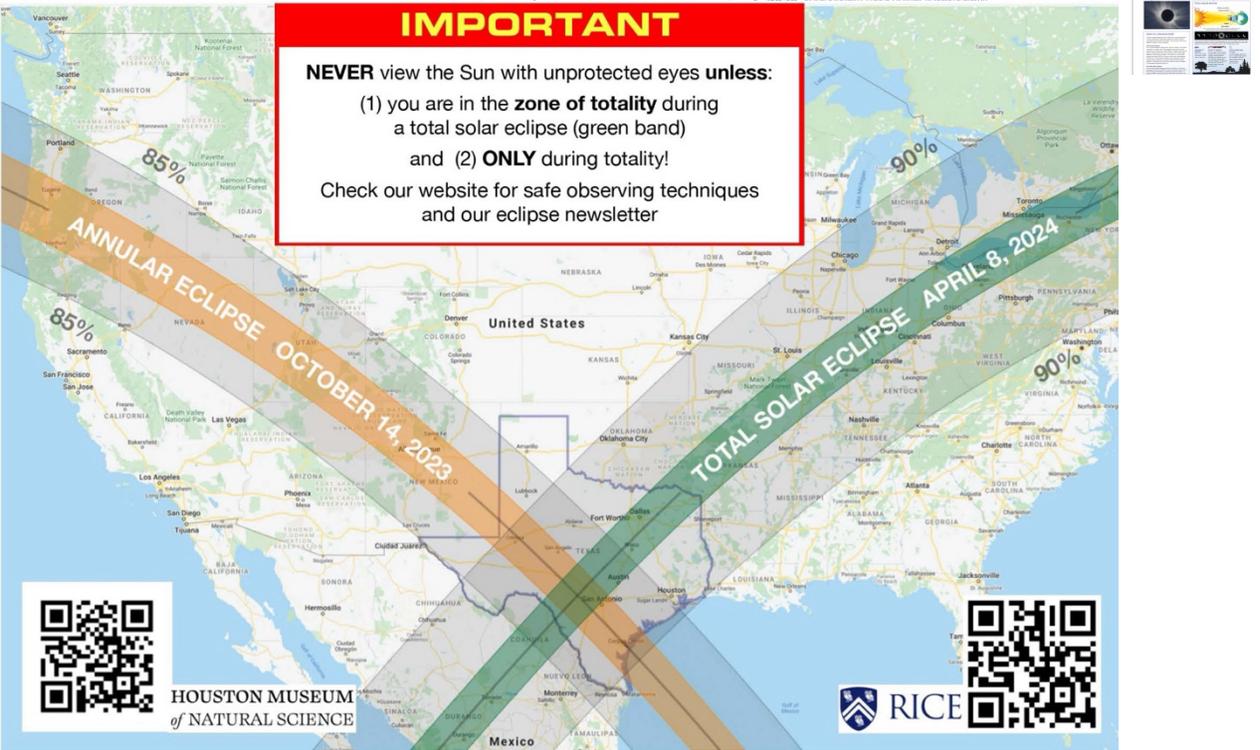

**Equipment:**

We also posted a list of our favorite eclipse observing equipment.  Since this did include specific recommendations for makes and models, we host this page on an external (non-Rice University) site: https://texaseclipse.net/equipment.html.  Binocular tripod adapters (which allow standard binoculars to be mounted on photographer tripods), slip-on solar filters for telescopes, binoculars, and telephoto lenses, 2× foldable eclipse binoculars, and gaffer tape are the most cost-effective additions to a solar viewing kit.

We designed both special "eclipse glasses" and "eclipse viewing cards," made for us by Rainbow Symphony.  We gave away more than 35,000 eclipse glasses and 30,000 eclipse viewing cards to teacher and public events.  In addition, Sumners and her intern Bethany Elliott designed a version of the eclipse viewing card paid for by Buc-ee's® and distributed by HMNS (Fig 11).  Each public school in Texas (over 8,800) could request 100 cards for their school; using lanyards, each card could be used by dozens of students. Thus, over a million eclipse viewers were distributed, reaching every county and school district in the state.

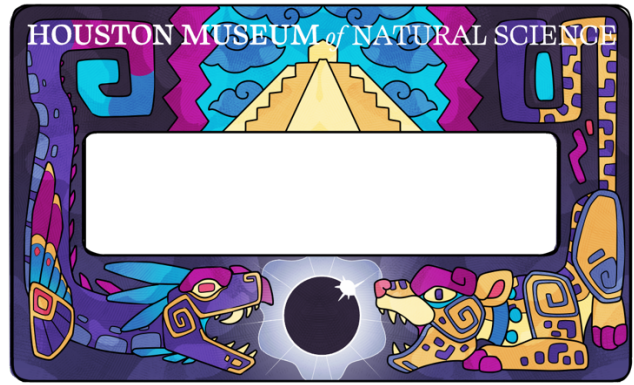

*Fig 11.  Special Mayan-themed eclipse viewing card, created by student artist Bethany Elliott*

**Annular Eclipse Public Events**

For the October 2023 annular eclipse, Reiff took a group of students and friends to observe on the beach near Corpus Christi (where the annular left the U.S.)  We tried out the "solar cup projector" on the eclipse (Figure 8), and also used a Sunspotter®, filtered binoculars and other devices.  We showed the crowd that gathered all the safe techniques.

Sumners similarly took a group to the "Texas Nexus," a place near Bandera which was in the path of both the annular and total solar eclipse.  There they also practiced the equipment they would use for the total eclipse. There was also a live public event on the plaza at the Houston Museum of Natural Science, which drew about 2500 people, observing through four telescopes and several carts, giving away solar viewers and demonstrating solar toys and a solar cooker.

Caspi, Kovac, and others on the Citizen CATE 2024 team established two high-altitude public observing sites – at Loveland Pass, CO and at Sandia Peak, NM – as part of a special coronagraphy and outreach project that also served as a beta-test of the CATE 2024 equipment and procedures (see below) prior to the April 2024 total eclipse. Outreach at these sites included public engagement and distribution of solar viewing glasses. Events at Loveland Pass were livestreamed on YouTube (https://www.youtube.com/watch?v=tHL1U74tvyo and https://www.youtube.com/watch?v=7ZWWZTIzWdM) and events at Sandia Peak were featured in an episode of PBS Nova (S51E6, "Great American Eclipse," air date April 3, 2024). Full details of these expeditions, including observations, can be found in Seaton *et al*. (2024).

**Citizen CATE 2024:**

The Citizen CATE 2024 project (https://eclipse.boulder.swri.edu/citizen-cate-2024/) was a major observing experiment for the TSE of April 8, 2024, led by SwRI and in partnership with NSF NCAR and other institutions, and funded by both NSF and NASA. CATE 2024 aimed both to observe the total solar eclipse for its entire duration over the continental U.S., and to provide scientific opportunity and engagement for interested parties in communities along the eclipse path. The scientific goal of CATE 2024 was to create a continuous movie of the corona during the approximately one hour of totality over the U.S., to study changes in the inner and middle corona that would be uniquely visible in CATE's high-cadence, long-duration observation. To achieve this, 35 sites were chosen such that totality would be visible by two sites at all times (Fig 12), in case one site was clouded out or had technical difficulties. Each of the 35 sites was operated by a team of non-astronomer (and generally non-scientist) volunteers from a nearby community, who were provided with identical observing equipment and procedures to obtain data in a uniform manner. Team participants included students (middle school, high school, undergraduate, and graduate), educators, non-scientist professionals, and retirees. Each site's equipment complement included an 80 mm refractor telescope (with solar filter for observing outside of totality), an equatorial tracking mount, and a digital CMOS camera which registered the polarization of the light in four different angles separated by 45°, thereby covering all polarizations. These polarization measurements enable unique visualizations [*Patel et al. 2023*] and provide some depth determination along the line of sight in the inner corona. The equipment then was donated to the community groups to serve as an educational and public outreach resource into the future.  In addition, eight other sites were distributed along the path using the same equipment setup; four were staffed by professionals, three were for outreach purposes only (no data collection), and one was a community volunteer team from the U.S. Air Force Academy.

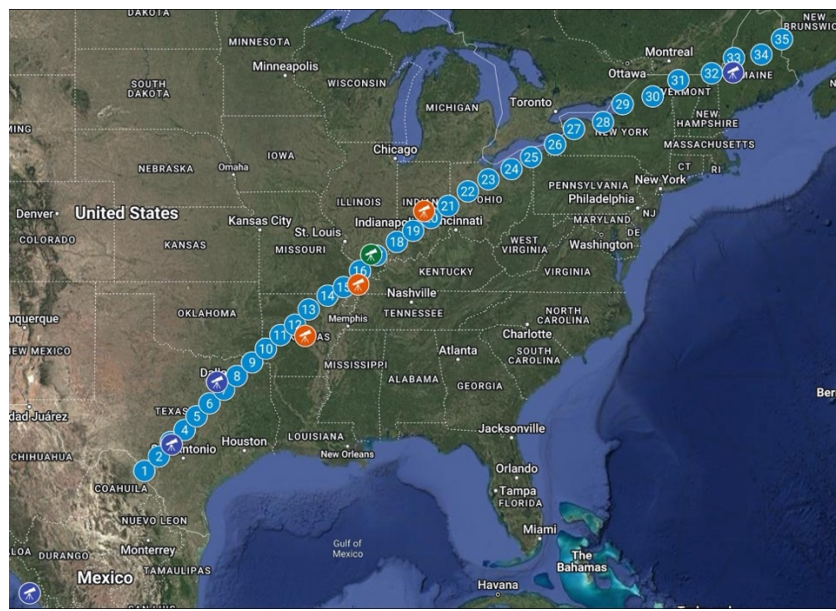

*Fig 12. CATE 2024 viewing sites. The 35 community-led sites are marked in light blue. Additional sites, led by professionals or outreach teams, are shown in orange or dark blue. Map image courtesy Google Maps.*

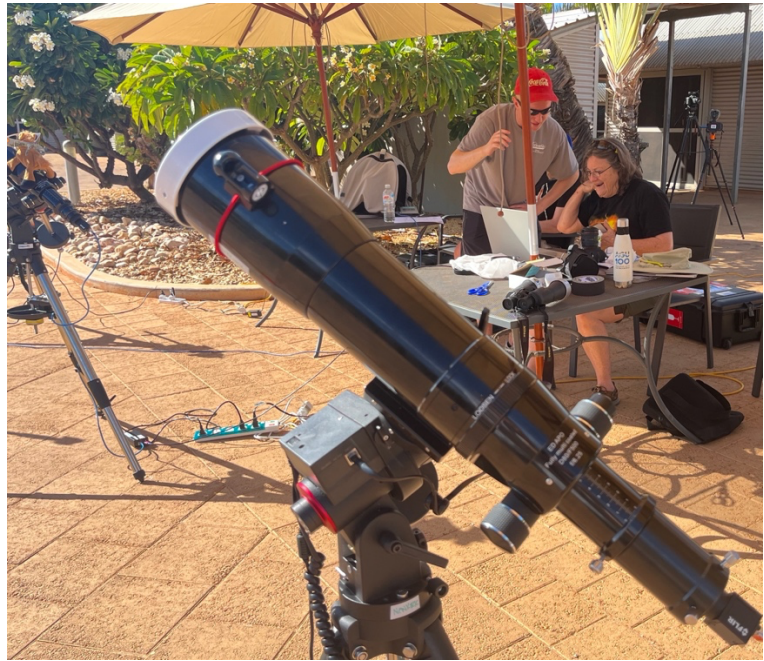

*Fig 13.  CATE telescope setup (foreground) after the Exmouth eclipse.  Charlie Gardner and Patricia Reiff are in the background analyzing the eclipse imagery.*

The CATE 2024 project – its conception, development, implementation, and first examples of its eclipse observations – are presented in detail in *Kovac et al.* (2025).  A 6-minute mini-feature describing the project is available online (https://www.youtube.com/watch?v=EAH9oqrfQc8). NSF NCAR produced a 26-minute documentary highlighting the community impact and participation of this project, "Gathered in Darkness," which was shown by various PBS stations along the eclipse path; a screening can be requested at https://edec.ucar.edu/public/field-projects/cate-2024.

Training for the volunteer teams was provided by SwRI through a network of coordinators.  The path was divided into three regions:  Southwest, Central, and Northeast, with approximately 12 sites in each.  Each region had a "regional coordinator" and a "lead trainer." Those regions were further divided into "State" groups of four to five teams, with a State Coordinator and their own lead trainer helping to find teams and then training and assisting each team.   The regional coordinators and their lead trainers were initially trained on the equipment at SwRI in Boulder, CO and then did a live test of the hardware and software at the total solar eclipse in Exmouth, Australia in April 2023 (*Patel et al.*, 2023; a livestream recording is available at https://www.youtube.com/watch?v=-jdW4BUptGo).  One of us (Reiff) was the Southwest

Regional coordinator, with lead trainer graduate student Charlie Gardner (Fig 13).  All of the regional and state coordinators, and their respective lead trainers, participated in a training workshop at SwRI in San Antonio, TX in January 2024, and State Coordinators then held their own workshops in subsequent weeks to train their local teams. Each team received their equipment at the workshop and then held multiple practices leading up to eclipse day, including three nationally-coordinated practices by all teams simultaneously, and a dry-run the day before the eclipse. Teams were able to share their experiences and lessons learned through a project-specific Slack workspace, which helped provide rapid dissemination of information and procedure improvements to enhance both the volunteer experience and the uniformity and quality of the scientific data.

On April 8, 2024, the day of the eclipse, every CATE 2024 team participated in the field, including all community-led, professional, and outreach teams. Every team participated fully and all of the equipment and procedures functioned appropriately, with no equipment failures or mishaps. A handful of teams were unfortunately clouded out, particularly in southwest Texas and around Lake Erie, but all teams attempted observations and over 80% of teams successfully observed some or all of totality. The earliest data were processed within 24 hours and were presented at the Triennial Earth-Sun Summit (now called Heliophysics Summit) in Dallas, TX just two days after the eclipse, and were also presented in a SwRI press release (https://www.swri.org/newsroom/press-releases/swri-led-eclipse-projects-shed-new-light-solar-corona). Full details of the observations, including statistics, are discussed in *Kovac et al*. (2025).

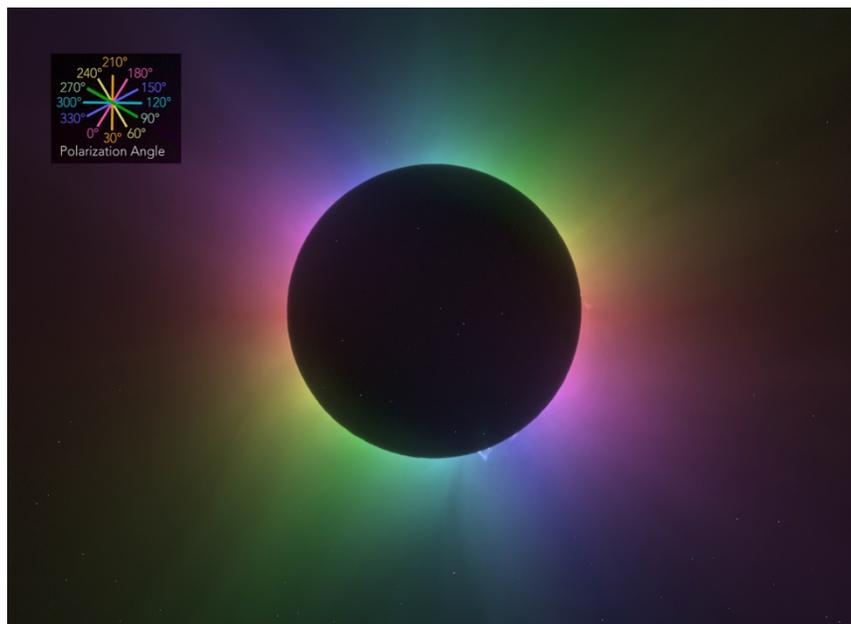

Each site's observations included a sequence of 8 images with varying exposure times that are combined to form a single "high dynamic range" (HDR) image. Each HDR sequence spans about 1.6 seconds, so each site acquired about 130–170 HDR images of totality. Each HDR image includes the full polarization data, which can be color-coded by polarization angle as shown in Figure 14.  The base image was taken by team 7 (Kemp ISD in partnership with Texas A&M) in Texas during the April 8 2024 total eclipse. The angle of polarization is indicated by the color, and the color saturation indicates the relative amount of polarization. Prominences, which are largely unpolarized, render as white (seen near the bottom). The polarized portion of the corona is caused by scattering of sunlight by free electrons orbiting the magnetic field lines, resulting in polarization perpendicular to the field lines. Since the field lines are nearly radial, the polarization is nearly tangent to the local surface, so the color advances around the spectrum twice around the coronal image (see inset for approximate color for each angle). Deviations from the radial direction contain important information about coronal conditions, and the data is also being processed in other ways (as discussed in Kovac et al. 2025) to then contribute to detailed scientific analysis. In total, more than 250 participants (including citizens, students, and amateur astronomers) participated in this effort. The analysis of these unique observations is still underway.

*Fig 14.  CATE 2024 coronal image from team 7 in north Texas, color coded by the direction of polarization of the coronal light.  Adapted from Team 7 HDR images.*

**Media Appearances and Eclipse Day observing**

As the eclipse approached, both Reiff and Sumners participated in many events for the media, including radio, print, and TV: https://space.rice.edu/eclipse/eclipse_media_coverage.html is a partial list. Reiff also did a briefing for media representatives, sponsored by SciLine, which resulted in several media appearances and a video piece distributed by the Associated Press. We also organized an eclipse expedition sponsored by the School of Natural Sciences at Rice, with more than 300 alumni, faculty and students present https://www.youtube.com/watch?v=ALhcWSI5I7s. Three TV stations (from Houston and San Antonio) came to film the eclipse live, but unfortunately the clouds did not cooperate.

In addition, the CATE 2024 team, including professionals and community volunteers, made many additional public and media appearances. Caspi, Kovac, and others were featured in a PBS Nova episode shown prior to the eclipse, including a feature on a unique CATE-affiliated project held during the 2023 annular eclipse. Caspi participated in a number of media interviews leading up to and including eclipse day, including for CBC, the SETI Institute, and others. On eclipse day, Caspi helped run a CATE site (led by Dan Seaton, the CATE 2024 Project Scientist) at the NASA-/NSF-/NOAA-sponsored Cotton Bowl event in Dallas, TX, while Kovac led a CATE outreach site at a NASA-sponsored event in Kerrville, TX.

Eclipse Day at HMNS also drew a packed crowd on the plaza, with more than 2500 people viewing through four telescopes and several carts passing out eclipse viewers and showing Sunspotters®, solar cookers and solar toys.

Total reach from all media outlets and public event sis unknown but is clearly in the tens of thousands.

**Space Weather forecasting**

With funding from the Magnetospheric Multiscale Mission (MMS), we post highly accurate but short-term forecasts of space weather (https://mms.rice.edu/forecast.html). The forecast uses IMF and solar wind data from the ACE and DSCOVR spacecraft at L1.  Our prediction uses the "Boyle Index" (Boyle et al., 1997) as the base input function and a neural net-derived weighting of that input over the previous 2-12 hours to predict the global response (Bala and Reiff, 2014). We create one-hour and three-hour ahead predictions of Kp (a global activity index, logarithmic with 0 indicating quiet and 9 being maximally disturbed); Dst (a measure of the equatorial magnetic field depression in nT, caused the enhanced ring current; and AE (a measure of the magnetic effects of the auroral electrojet, measured in nT).

We send "Yellow" alerts to our >900 "Spacalrt" subscribers if Kp is predicted or measured to be over 4, and Red alerts if that measure exceeds 6.  Typically, most Red alert conditions, which are major geomagnetic storms, occur near or just after solar maximum.

In May 2024, a giant CME passed by the solar wind monitors at L1, which resulted in our Kp forecast up to Kp9, (Fig 15).   Accordingly, we sent out email alerts not only to our Spacalrt subscribers, but also notices to all our outreach lists (over 20,000 subscribers) plus posted to Meta and X. Sure enough, auroras were seen in California, Colorado, and even Texas.  A number of people sent us thanks and photographs of an amazing event they would have missed out of without the reminder to "GO OUTSIDE NOW!" Our predictions for Kp were amazingly accurate (Fig 13), and we even correctly predicted the maximum Dst (-400) and AE (2000) indices.

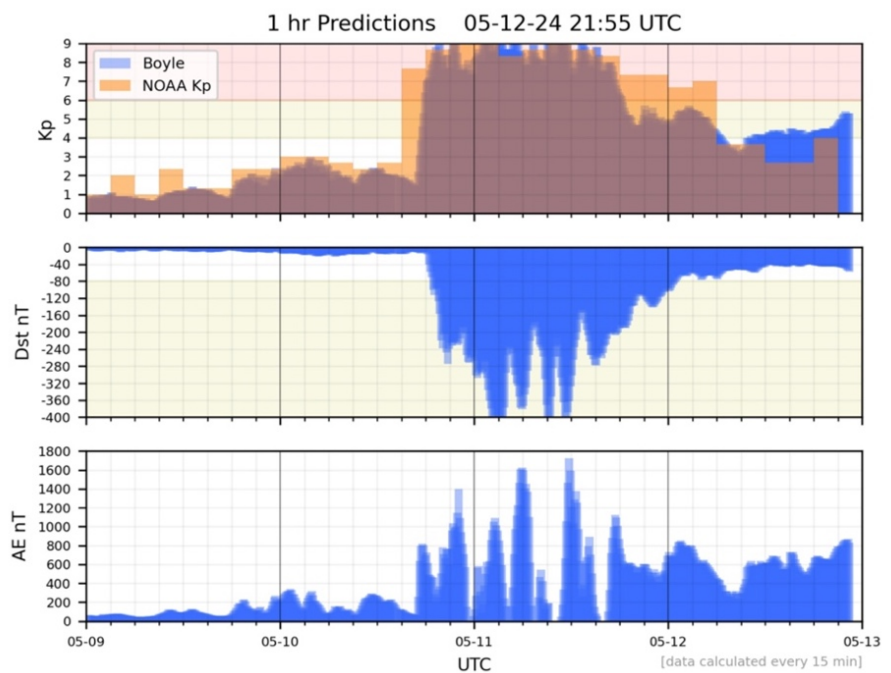

*Fig 15.  Our one-hour-ahead predictions for the Kp index (top, blue), along with the NOAA real-time estimate (orange), for the May 2024 "Gannon Storm".  The middle panel is the predicted Dst index (nT), and the lower panel is the predicted AE index (nT).*

The October event, for which we also sent out a Red alert, also triggered some low-latitude auroras, but not nearly to the extent of the May storm. One key to visibility of the aurora for North American viewers is of course the Universal Time of the storm maximum – if it occurs between about 0–7 UT, its auroras (which are strongest in the evening through midnight) will be visible in the Americas.

## 4. Results:

For the "Heliophysics Big Year," the partnership of Rice University and the Houston Museum of Natural Science reached well over 250,000 documented views in leveraged venues (teachers, museums, planetariums), plus uncounted visitors to websites and media appearances. We partnered with events from NASA, ESIP Federation, Hofstra University, SEEC, NSTA and CAST. We participated in and helped coordinate the Citizen CATE 2024 research program and helped advise SCIHam efforts. Also not listed here were the dozens of public presentations by the authors at Rice, HMNS, Science Café, Astronomy on the National Mall, Scouts, etc. and media appearances in print, radio, web and TV. We distributed eclipse viewers to almost a million teachers. Our total reach was well over 1.5 million people. We confirmed 5405 teacher users of our materials. The user survey showed the materials we developed were of significant benefit and were widely disseminated by the teachers, on average reaching 500 students each.

## 5. Discussion:

The substantial investment in federal funds, planetarium, university and research institute resources and even private corporations combined to make the Total Solar Eclipse of 2024 probably the best observed eclipse in human history. Our project used all these resources, and also brought in and trained dedicated amateur and student observers to take crucial observations in the CATE 2024 program. Together we provided an important source of eclipse information used in the US and around the world and the results will have lasting impact.

**Acknowledgements:**

Partial funding for animations used in the "Totality" planetarium show was from the NASA "HEAT" project at Rice University grant 80NSSC21K1563.  Additional funding for completion of the planetarium show was provided by the Houston Museum of Natural Science and by MTPE, Inc.  (*https://www.ePlanetarium.com*).   Funding for additional teacher trainings and public events were provided by the NASA MMS Mission at Rice University under grant 599790Q through Southwest Research Institute.  Funding for nearly a million eclipse viewers was provided by Buc'ees, Inc.  Funding for Citizen CATE 2024 was provided by grants from the National Science Foundation (award numbers 2231658, 2308305, 2308306, 2511904, 2511905, & 2511907), NASA (grant numbers 80NSSC21K0798 & 80NSSC23K0946), and SwRI (internal award R6395).

**Appendix:**

**Websites and web links**

https://space.rice.edu   Overarching space outreach site
https://space.rice.edu/eclipse  Eclipse information sites
https://mms.rice.edu/forecast.html  MMS space weather forecasting site

https://mst.rice.edu/ASTR502 ASTR 502 Solar System class for Teachers
https://mst.rice.edu/ASTR503,  ASTR 503, Astronomy for Teachers
https://mst.rice.edu/ASTR530, ASTR 530 Teaching Astronomy Lab
https://mst.rice.edu/PHYS501 PHYS 501 Physics of Ham Radio class

https://fireflyspace.com/news/blue-ghost-mission-1-live-updates/ Firefly images of the lunar eclipse

https://eplanetarium.com/shows/ddome/hmns/totality/  "Totality" page at ePlanetarium

https://space.rice.edu/eclipse/eclipse_show.html  "Totality" page at Rice University

https://eclipse.boulder.swri.edu/citizen-cate-2024/  Citizen CATE 2024 site

https://www.youtube.com/watch?v=8zbZnZOW0Rw "Gathered in Darkness" video about CATE.